# Design and development of surface modified p and n type silicon sensor for nitrogen gas flow measurement

U. Satheesh, P. Prakash and D. Devaprakasam*

*NEMS/MEMS and Nanofabrication Lab (InCUBE), Department of Nanosciences and Technology,
Karunya University, Coimbatore – 641114, India.*

**Abstract**

We report a gas flow driven voltage generation of Octyl trichlorosilane (OTS) molecules self assembled on silicon wafers (Si wafers). OTS Self assembled Monolayer (SAM) has been coated on both p-type and n-type doped silicon wafers (p-Si and n-Si wafers) using dip coating method. We have measured the flow induced voltage generation on OTS SAM coated Si wafers/Uncoated Si wafers at modest gas flow velocities of subsonic regime (Mach number < 0.2) using national instruments NI-PXI-1044 Workstation. The gas flow driven voltage generation is mainly due to the interplay mechanisms of Bernoulli's principle and Seebeck effect. The surface morphology of OTS SAM coated p-Si and n-Si wafers were characterized by SEM analysis. In this study, our results shows that OTS SAM coated p-Si and n-Si wafers shows better sensitivity towards nitrogen gas flow when compared with the uncoated Si wafers. OTS SAM also exhibits high thermal stability and hydrophobicity.

## 1. Introduction

The recent studies on flow driven voltage generation shows that a measurable amount of electrical signal can be generated across the solid surface, when gas/liquid flows over a surface of the solids like CNT, Graphite and doped semiconductors at modest velocities [1-5]. In our previous work, we have reported that when a nitrogen gas flows over an inclined surface at an angle of $\pi/4$ of the surface modified p-type and n-type doped semiconductors, it could generate a finite electrical signal due to the interplay mechanisms of Bernoulli's principle and Seebeck effect. We have also reported the in-house design and development of a subsonic regime flow measurement experimental system. Our objective of the study is to measure the generated electrical signal due to the temperature gradient and to optimize the generated electrical signal by modifying the surface of the p-Si and n-Si wafers with the self assembled monolayers. By using different methods like liquid phase [6] and vapour phase [7-10], SAM can be easily patterned on Si wafers. The perfluoro alkylsilane SAM like FOTS, OTS, FDTS, FOTES, FOMMS and FOMDS are highly hydrophobic and have high thermal stability [6, 9-11].

In this work, we have studied the nitrogen gas flow driven voltage generation of Octyl trichlorosilane self assembled on both p-type and n-type doped Si wafers. Our results shows that self assembly molecules of OTS coated p-Si and n-Si wafers shows better sensitivity response than uncoated Si wafers due to high thermal stability and hydrophobicity. We also report that OTS SAM coated p-Si and n-Si wafers have sensitivity towards nitrogen gas flow and convert the gas flow energy into finite electrical signal. The OTS SAM coated p-Si and n-Si wafers could be used as a flow sensor and finds applications in energy harvesting devices and gas flow sensors.

## 2. Experimental Details

### 2.1 Material Details

P-type silicon wafer (single side polished) <100> contains boron as dopant 0.5 mm thickness, N-type silicon wafer (single side polished) <100> contains phosphorus as dopant 0.5 mm thickness, Octyl trichlorosilane (OTS) (97% Purity), Isooctane (99.8% Purity) were all commercially purchased from Sigma Aldrich, India. Conductive silver paste was purchased from Siltech Corporation Inc, India.

---

* Corresponding author. Tel.: +91-422-2164488    fax: +91-422-614614
  *E-mail address*: devaprakasam@karunya.edu

*2.2 Preparation of OTS SAM*

1mM mixture of OTS is prepared in isooctane solvent. The Si wafers of 18x18 mm$^2$ were cut from p-type (100) and n-type (100) Si wafers. The p-Si and n-Si wafers were immediately immersed in a freshly prepared ultrasonicated solution with its polished side facing up for duration of about 30 mins. After SAM deposition, the p-Si and n-Si wafers were taken out and allowed to dry in a room temperature condition. The OTS SAM formations on the Si wafers are shown schematically on figure.1. Thus the OTS SAM coated p-Si and n-Si wafers could be used as a sensor for flow measurement.

*2.3 Experimental Setup*

The OTS SAM coated and uncoated p-Si and n-Si wafers are attached over the aluminium mount individually. But the Si wafers and the aluminium mount are separated thermally and electrically by using insulating materials. The two copper connection leads are taken from the Si wafers using silver conductive paste. The aluminum mount along with the attached Si wafer is placed over the subsonic regime flow measurement experimental setup and the connection leads are connected to the National instruments NI-PXI-1044 Workstation.

*2.4 Characterization Techniques*

After OTS SAM deposition on p-Si and n-Si wafers, the surface morphology studies were characterized by SEM analysis. The figure.2 shows the schematic of the experimental setup. The designed and developed flow sensor is connected to the national instruments NI-PXI-1044 Workstation using connection leads and we have measured the current and voltage response of p-Si and n-Si wafers at various gas flow velocities.

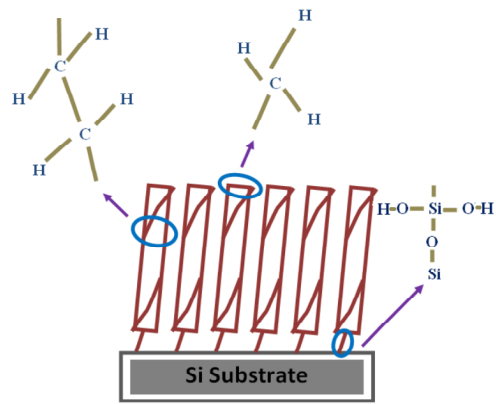

Figure.1. Schematic of OTS SAM coated p-type and n-type Si wafers

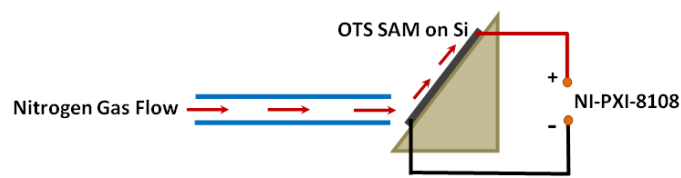

Figure.2. Schematic of the experimental setup for nitrogen gas flow measurement

## 3. Results and Discussion

*3.1 Morphology Analysis*

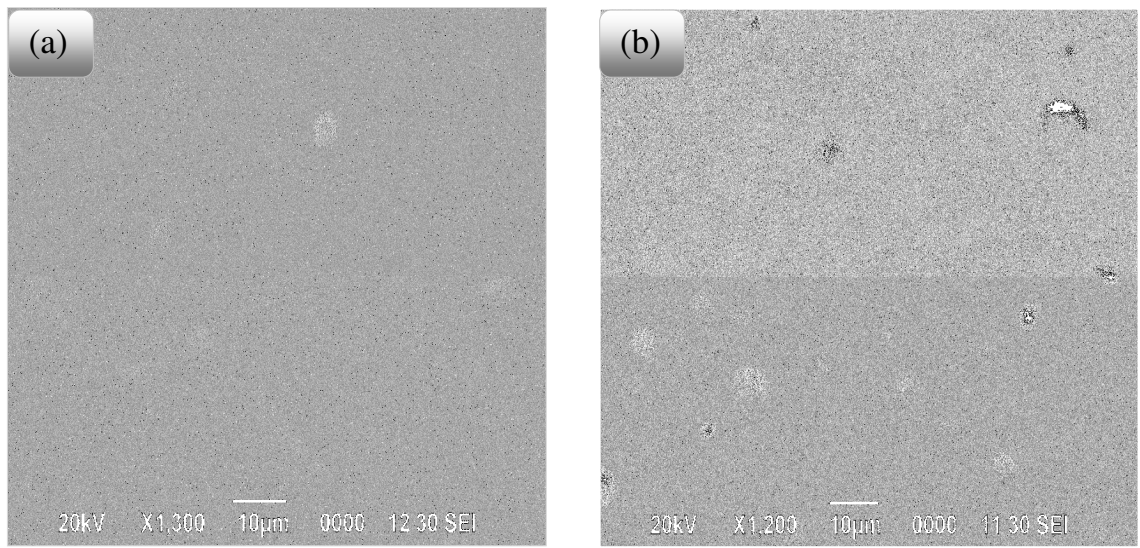

Figure.3. (a) and (b) shows the SEM images of OTS SAM coated p-type and n-type Si wafers

    The OTS SAM has been coated on the p-Si and n-Si wafers by dip coating method for about 30 mins. After the SAM coating process, the OTS SAM coated Si wafers were allowed to dry under the room temperature condition. The surface morphology of OTS SAM coated Si wafers were characterized by SEM analysis. The figure.3 shows the SEM images of OTS SAM coated p-Si and n-Si wafers.

*3.2 I-V Measurements*

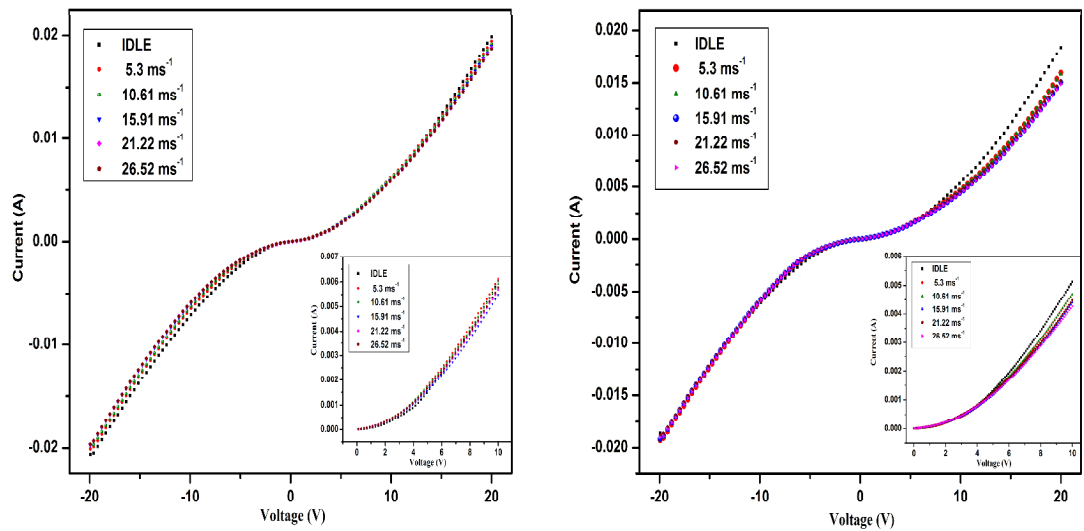

Figure.4. Shows the (a) I-V curve of the uncoated p-Si wafer (b) I-V curve of the OTS SAM coated p-Si wafer.

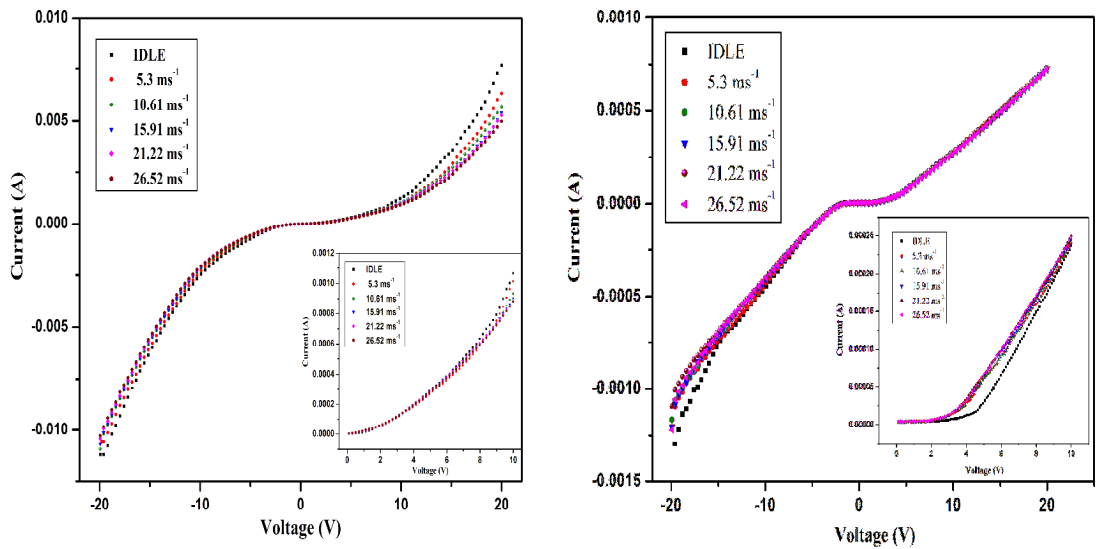

Figure.5. Shows the (a) I-V curve of the uncoated n-Si wafer (b) I-V curve of the OTS SAM coated n-Si wafer.

I-V measurements were carried out on the uncoated and OTS SAM coated p-Si and n-Si wafers at different gas flow velocities in the range of 5.3 to 26.52 ms$^{-1}$ using National instruments NI-PXI-8108 workstation. Here we have used the nitrogen gas as a gas flow and we have studied the flow induced voltage measurement. The nitrogen gas from the compressed gas cylinder is allowed to flow over the surfaces of uncoated and OTS SAM coated p-Si and n-Si wafers at a given pressure (maximum pressure of 150 bars) through the flow meter and flow tube. The flow meter connected to the nitrogen gas cylinder measures the flow rate of the nitrogen gas and the flow rate can be controlled using the valve in the flow meter. We have measured the I-V curve for a voltage range of between -20 V to 20 V for both uncoated and OTS SAM coated p-Si and n-Si wafers. The figure.4 (a) and (b) shows the I-V curve for the uncoated and OTS SAM coated p-Si wafers. The figure.5 (a) and (b) shows the I-V curve for the uncoated and OTS SAM coated n-Si wafers. The inset present in each graph shows the corresponding I-V curve for a voltage range of between 0 V to 10 V.

*3.3 Sensor Voltage Measurements*

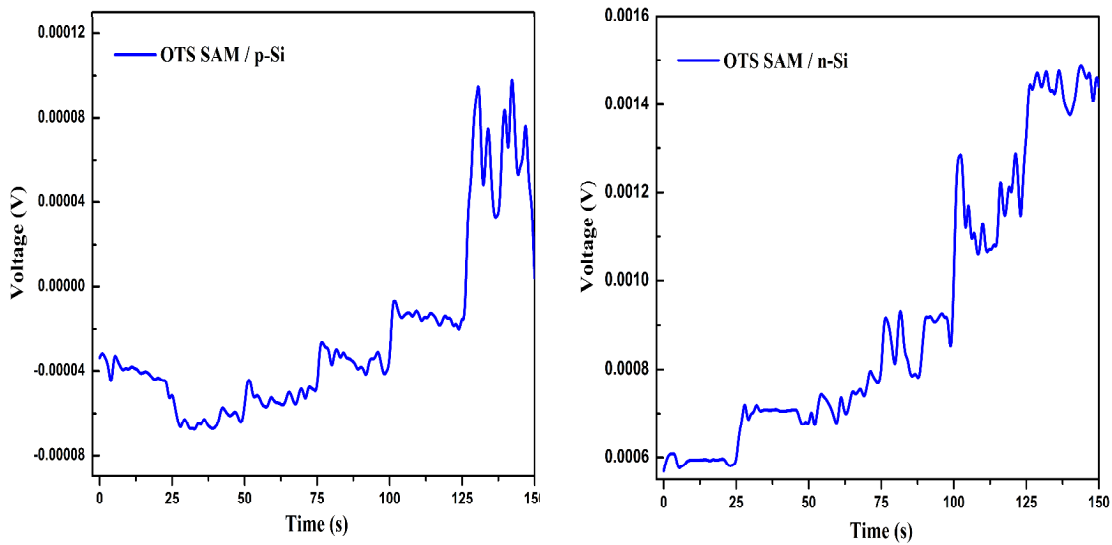

Figure.6. (a) and (b) shows the response voltage (V) at given time interval (s) with the dependence of nitrogen gas flow velocity for the OTS SAM coated p-Si and n-Si wafer.

We have measured the response voltage (V) for the OTS SAM coated p-Si and n-Si wafers at a given time interval (s) for various gas flow velocity in the range of 5.3 ms$^{-1}$ to 26.52 ms$^{-1}$ using National instruments NI-PXI-8108 workstation. The response voltage was measured at a given time interval of 25 seconds each for no flow and at various nitrogen gas flow velocities like 5.3, 10.61, 15.91, 21.22 and 26.52 ms$^{-1}$. Figure.6. (a) and (b) shows the response voltage (V) at given time interval (s) with the dependence of nitrogen gas flow velocity for the OTS SAM coated p-Si and n-Si wafers. From the results, we observed that the OTS SAM coated p-Si and n-Si wafers could be used as a flow sensor towards the nitrogen gas flow and in turn generates a finite electrical voltage. Thus the OTS SAM coated p-Si and n-Si wafer exhibits a voltage of 0.08 mV and 1.4 mV for the nitrogen gas flow velocity of about 26.52 ms$^{-1}$. Thus the generated electrical voltage can be further increased by increasing the flow velocity of the nitrogen gas over the OTS SAM coated p-Si and n-Si wafers.

## 4. Conclusions

In this work, we have designed and developed a gas flow sensor by modifying the surfaces of the p-type and n-type silicon wafers with OTS self assembled monolayer. The flow sensor works based on the interplay mechanism of Bernoulli's principle and Seebeck effect. We have studied the flow induced voltage generation at modest velocities of subsonic regime, where Mach number < 0.2. We have measured the current and voltage response of p-Si and n-Si wafers at various gas flow velocities. At 26.52 ms$^{-1}$ velocity, OTS SAM coated p-type and n-type Si wafers generated 0.08 mV and 1.4 mV respectively. From the results, we conclude that OTS SAM coated p-Si and n-Si wafers could be used as a flow sensor towards the nitrogen gas flow and generates a measurable finite electrical voltage. But, however the generated electrical voltage can be further increased by increasing the flow velocity of the nitrogen gas. Our previous work shows that the OTS SAM exhibits high thermal stability and hydrophobicity. The designed and developed sensor finds applications in energy harvesting devices and gas flow sensor.

**Acknowledgements**

We thank the DST-Nanomission, Govt. of India and Karunya University, Coimbatore for providing necessary financial support, to carry out this research work. We also thank the Department of Nanosciences and Technology, Karunya University for their help and support during the research work. We thank A. Raja and M.B.S. Praveen, lab technicians for their support.